\documentclass[final]{svjour3}
\usepackage{graphicx}
\usepackage{rotating}
\usepackage{amsmath}
\usepackage{amssymb}
\usepackage{mathptmx}
\usepackage[numbers]{natbib}
\makeatletter
\journalname{Journal of Low Temperature Physics}

\bibpunct{}{}{,}{s}{}{,}

\begin{document}

\newcommand{\hdblarrow}{H\makebox[0.9ex][l]{$\downdownarrows$}-}
\title{Superconductor-ferromagnet tunnel junction thermoelectric bolometer and calorimeter with a SQUID readout}

\author{Z. Geng \and A. P. Helenius \and T. T. Heikkil{\"a} \and I. J. Maasilta}

\institute{Nanoscience Center, Department of Physics, University of Jyv{\"a}skyl{\"a},\\ Jyv{\"a}skyl{\"a}, FI-40014, Finland\\ Tel.: \\ Fax:\\
\email{zhgeng@jyu.fi}}

\maketitle

\begin{abstract}
The superconductor-ferromagnet thermoelectric detector (SFTED) is a novel ultrasensitive radiation detector based on the giant thermoelectric effect in superconductor-ferromagnet tunnel junctions. This type of detector can be operated without the need of additional bias lines, and is predicted to provide a performance rivaling transition-edge sensors and kinetic inductance detectors. Here we report our numerical studies on the SFTED noise equivalent power, energy resolution and time constant, and the feasibility of a SQUID readout in both bolometric and calorimetric regimes, with the goal to provide practical design parameters for the detector fabrication and the readout circuitry implementation.


\end{abstract}

\section{Introduction}

The superconductor-ferromagnet thermoelectric detector (SFTED) is a novel low-temperature radiation detector\cite{Chakraborty2018,Heikkila2018} based on the recent discovery of the giant thermoelectric effect in superconducting-ferromagnet hybrids\cite{Ozaeta2014,kolenda}. In contrast to other commonly used ultrasensitive detectors such as the transition-edge sensor (TES)\cite{Ullom2015}, the kinetic inductance detector (KID)\cite{Grossman1991} or the superconducting tunnel junction (STJ)\cite{Kurakado1982}
for which the signals come from changes to a quiescent current or voltage,
SFTED directly utilizes the measurable electrical signal transduced from the radiation absorption without bias power, and therefore fundamentally cuts down the heat dissipation and wiring complexity for large sensor arrays. This feature can be extremely attractive for modern bolometer and calorimeter applications, in which large sensor arrays are preferred\cite{Ullom2015,Pirro2017}.

Here, we discuss the SFTED in both bolometric and calorimetric regimes. Numerical results of the detector performance, i.e. noise equivalent power (NEP), thermal time constant ($\tau_{th}$), and energy resolution ($\Delta E$), will be presented with realistic design parameters. We also explore the feasibility of using a dc Superconducting QUantum Interference Device (SQUID) for the readout. 
Using a SQUID has the added benefit that several well-developed multiplexing schemes for large arrays already exist \cite{Ullom2015,Kiviranta2003,Irwin2004}.

\section{Bolometer study}
The sensing element of a SFTED is a tunnel junction between a superconductor and a normal metal, where the superconducting density of states has been spin split by an exchange field  ($h$)  induced by a nearby ferromagnetic insulator. To generate the electron-hole asymmetry required for thermoelectric response, where a temperature difference $\Delta T$  between the electrodes produces a thermovoltage $V_{th}$  and thermocurrent $I_{th}$\cite{Ozaeta2014}, either the normal metal electrode or the tunnel barrier insulator has to be ferromagnetic (Fig. 1 (a)-(b)). Within the small signal regime, the thermal and electrical balance of the device can be expressed as:

\begin{equation}\label{eq:heat_eq}
\begin{aligned}
    C_h\frac{d\Delta T}{dt} &= P_{in}-G_{th}^{tot}\Delta T+\alpha V_{th} \\
    I_{th} &= \alpha\frac{\Delta T}{T}-GV_{th} \\
\end{aligned}
\end{equation}
where $C_h$ is the heat capacity, $P_{in}$ the absorbed power, $\alpha$ the thermoelectric coefficient \cite{Ozaeta2014}, and $G$ the junction dynamic electrical conductance.
The simplest relevant circuit that connects to the junction has a capacitance $C$ and an inductance $L$ in parallel, in which case the thermocurrent and thermovoltage are related by $I_{th}(\omega)=(i\omega C+1/i\omega L)V_{th}(\omega)$ at a particular frequency $\omega$.
In the thermal model used here (Fig. 1 (c)), the thermal conductance through the tunnel barrier $G_{th}$ and through the electron-phonon coupling in the superconductor $G_{e-ph}$ are assumed to be much smaller than the conductances to the phonon bath, thus $G_{th}^{tot}=G_{th}+G_{e-ph}$ is the total thermal conductance that dominates the heat flow.

\begin{figure}[htbp]
    \begin{center}
        \includegraphics[width=1\linewidth, keepaspectratio]{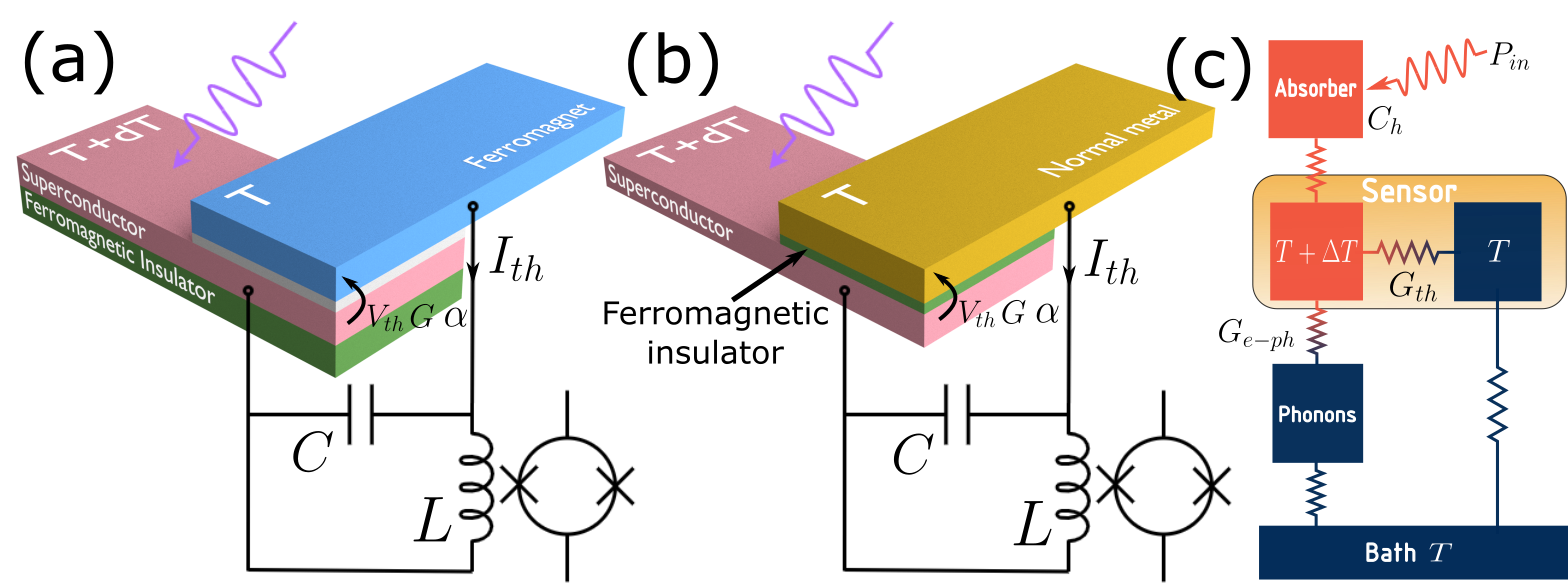}
        \caption{(a) and (b) Schematics of SFTED with SQUID readout. Superconducting electrode is used to absorb incoming radiation, and the temperature difference between the two electrodes drives the thermoelectric current $I_{th}$ tunneling across an insulating barrier made of either (a) AlOx or (b) EuS. Current signal will be coupled by an input inductor into a SQUID readout. (c) Thermal model of SFTED.}
        \label{fig:fig0}
    \end{center}
\end{figure}

In bolometric applications, NEP is often quoted as a benchmark for detector performance, which is defined as the input radiation power in 1 Hz bandwidth required by the detector to generate a signal equal to its noise. For a SFTED, based on Eq. (1) and the fluctuation-dissipation theorem\cite{Heikkila2018}, the detector will exhibit the typical Johnson noise and thermodynamic fluctuation noise (TFN), and in addition, a negative cross-correlation term between the junction current and heat current fluctuations, due to the strong thermoelectric response. The NEP for SFTED can then be derived as:

\begin{equation}\label{eq:NEP0}
NEP^2=\frac{4k_BT^2G_{th}^{tot}}{ZT}\left[1+(1+ZT)\tau_{th}^2\omega^2\right],
\end{equation}
where $\tau_{th}=C_h/G_{th}^{tot}$ is the thermal time constant and $ZT$ is the thermoelectric figure of merit \cite{Heikkila2018}. As one can see, a ZT value larger than unity would mean an improvement over a standard bolometer, made possible by the direct negative electrothermal effect on the noise. This improvement of NEP is seen to take place only at low frequencies below the effective thermal time constant $\tau_{eff}=\tau_{th}\sqrt{1+ZT}$, which also increases with $ZT$.

\begin{figure}[htbp]
\begin{center}
\includegraphics[width=0.8\linewidth, keepaspectratio]{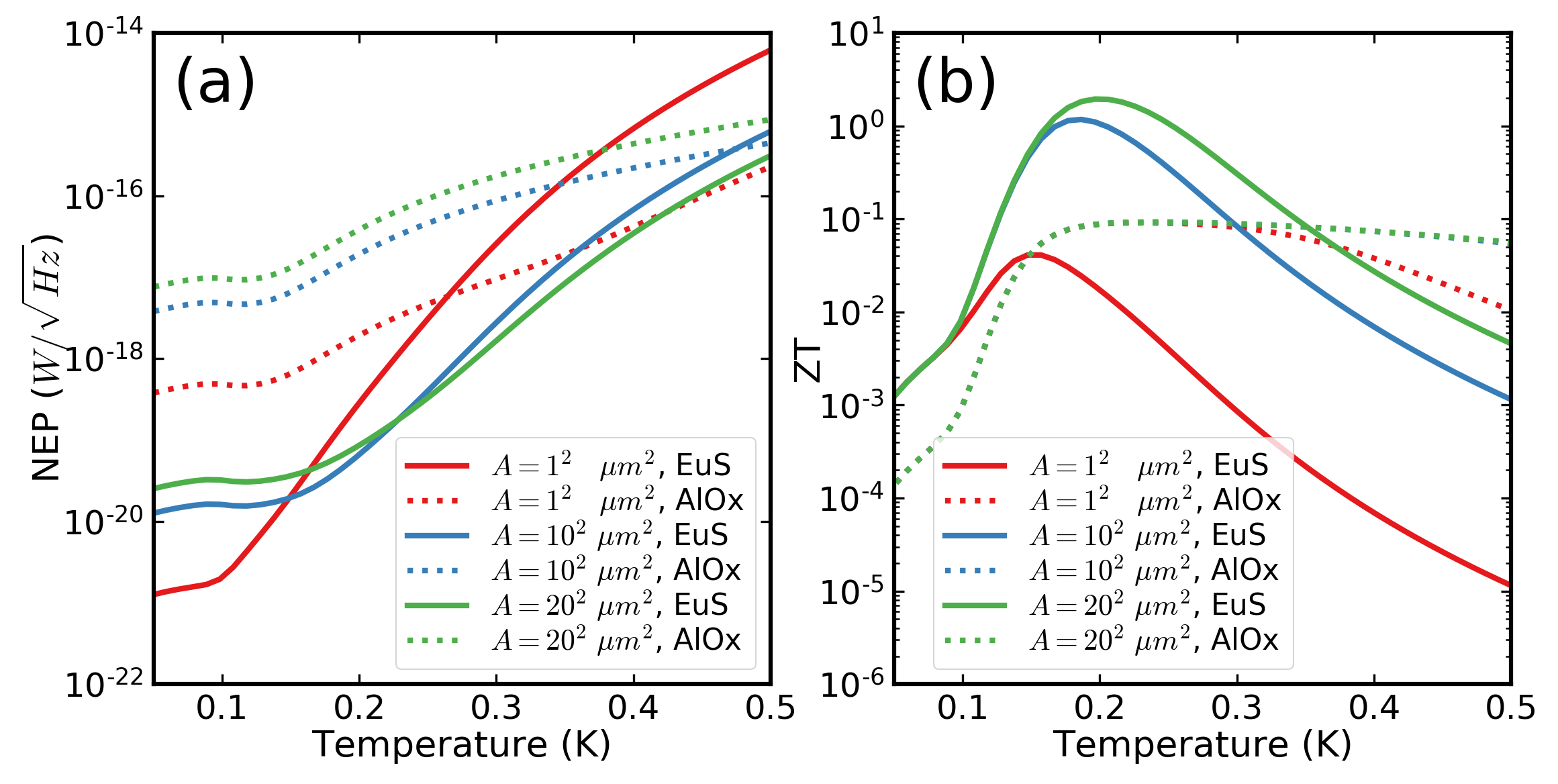}
\caption{Numerical calculation of zero-frequency NEP (a) and $ZT$ (b) of SFTED in the bolometric regime. {\it Color} indicates different junction areas, {\it solid lines} are junctions with EuS tunnel barriers ($P=0.9$), and {\it dotted lines} are those with AlO$_x$ barriers and ferromagnetic electrodes ($P=0.3$). In these plots, the superconducting electrode was assumed to be absorber and the material was Al with a volume of $V=2\ \mu m^3$, a broadening parameter of $\Gamma=10^{-4}\Delta$, and an exchange field of $h=0.3\Delta$. The specific junction resistivities used were $1\,k\Omega \mu$m$^2$ (AlO$_x$) and $10\,M\Omega \mu$m$^2$ (EuS). These calculations follow reference \cite{Chakraborty2018}, and take into account the modification of the superconductor energy gap and density of states due to the exchange field \cite{Giazotto2015}.}
\label{fig:fig1}
\end{center}
\end{figure}

Numerical calculations of the zero-frequency NEP (Fig.~\ref{fig:fig1}(a)) and the corresponding $ZT$ (Fig.~\ref{fig:fig1}(b)) for different junction sizes in a range 1 - 400 $\mu$m$^2$ are  presented as a function of operation temperature in Fig. ~\ref{fig:fig1}. Two different tunnel barrier materials, EuS and AlO$_x$ have also been compared in these plots.

EuS is a ferromagnetic insulator, and has been considered as a promising tunnel barrier material for SFTED in bolometric applications\cite{Heikkila2018}. When contacted with a superconductor, it induces a spin-splitting exchange field ($h$). At the same time, EuS can also function as a spin-filter between the superconductor and a normal metal electrode, to generate thermoelectric signals with a large polarization factor ($P$) exceeding 0.9\cite{Moodera2007}. As shown in Fig.~\ref{fig:fig1}, junctions with EuS barriers can have $ZT$ larger than unity, and are predicted to have a NEP below $100\ zW/\sqrt{Hz}$ with $\tau_{th}=0.1\sim40\ ms$ (not shown), improving over many previously reported detectors\cite{Karasik2011,Suzuki2014,DeVisser2014} for microwave and far-infrared applications. 

However, a ferromagnetic EuS barrier with a high polarization typically has a high tunneling resistance \cite{Moodera1988} ($10\,k\sim10\,M\Omega$ for the junction sizes considered here), thus a current sensing scheme based on a SQUID can be very hard to achieve. For a two stage SQUID readout with a low current noise $ \sim 100\,fA/\sqrt{Hz}$, the corresponding amplifier NEP is above $10\ aW$ and would thus dominate over the noise of the detector. To be matched with an SFTED, an on-chip large winding-ratio superconducting flux transformer with a gain $>500$ would be required. 

\section{Calorimeter study}

For calorimetric applications, the energy resolution $\Delta E$ (rms) is a figure of merit.
By applying optimal filtering\cite{McCammon2005} and assuming an infinite bandwidth amplifier, the energy resolution of SFTED can be obtained as\cite{Chakraborty2018}:
\begin{equation}\label{eq:dE}
\Delta E_{tot} = NEP_{tot}\sqrt{\tau_{eff}^{tot}},
\end{equation}
where $NEP_{tot}$ is the total zero-frequency NEP contributed by both the detector and the amplifier, and $\tau_{eff}^{tot}$ is the effective thermal time constant \cite{Chakraborty2018}.

\begin{figure}[htbp]
    \begin{center}
        \includegraphics[width=0.8\linewidth, keepaspectratio]{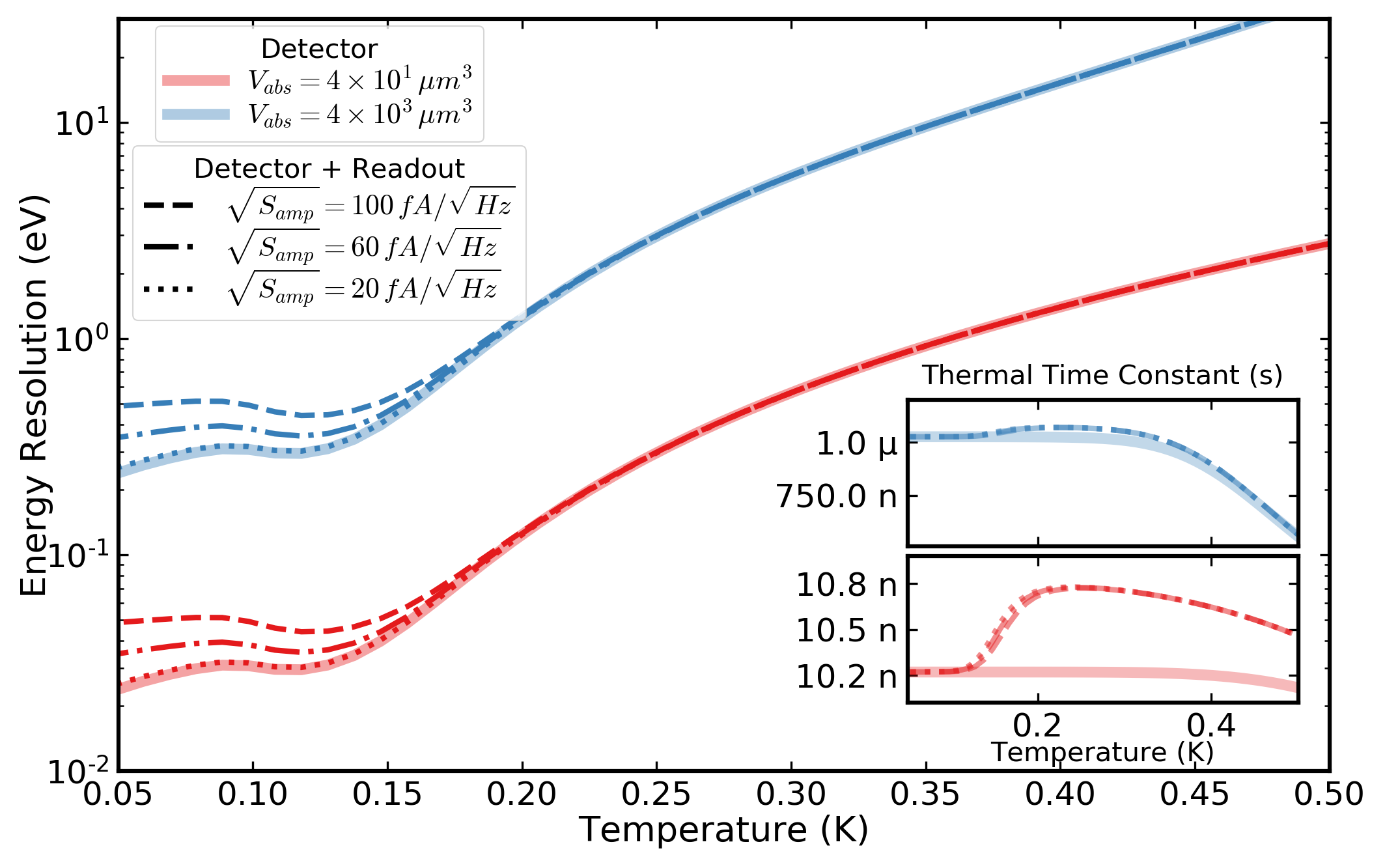}
        \caption{Numerical calculation of energy resolution ({\it main}) and thermal time constant ({\it insets}) of SFTED as a calorimeter. {\it Colored lines} indicate different absorber volumes. In both {\it main} and {\it inserts}, {\it solid lines} are the results for the detector alone, and {\it dotted, dashed} and {\it dash-dotted lines} are the results with both the detector and the readout with different SQUID current noise $\sqrt{S_{amp}}$. In these plots, the junction barriers were assumed to be AlO$_x$ with an area of $10^4\ \mu m^2$, a broadening parameter of $\Gamma=10^{-4}\Delta$, and an exchange field of $h=0.3\Delta$. The polarization factor used in calculation was $P=0.3$. For ferromagnetic electrodes such as Co or Fe, $P$ can be as high as $0.35\sim 0.4$ \cite{Meservey1994}.
        }
        \label{fig:fig2}
    \end{center}
\end{figure}

Calorimeters are routinely used for higher energy photon detection. To absorb the incident photon with high efficiency, an absorber structure with a larger volume compared to a bolometer is required. For SFTED at a low temperature, the heat transported through tunneling can be much larger than through electron-phonon coupling ($G_{th}\gg G_{e-ph}$). This fundamentally relaxes the requirement of a dedicated thermal isolation platform\cite{Ullom2015} such as a membrane, therefore easing the fabrication process. However, this also limits the use of high $P$ EuS as tunnel barrier for SFTED, since its lower specific transparency\cite{Moodera1988} will significantly increase the time constant and the amount of heat leaked through electron-phonon coupling. On the other hand, detectors with AlO$_x$ barriers can provide a very promising performance for calorimetric applications due to much higher transparency\cite{Greibe2011}.

A numerical calculation of energy resolution ({\it main plots}) and thermal time constant ({\it insets}) for SFTEDs with two different absorber volumes is presented in Fig.~\ref{fig:fig2}. In these plots, the tunnel barrier was AlO$_x$ and the absorber material was assumed to be Al for simplicity, and the volumes roughly correspond to X-ray detection.  

\begin{figure}[htbp]
    \begin{center}
        \includegraphics[width=0.8\linewidth, keepaspectratio]{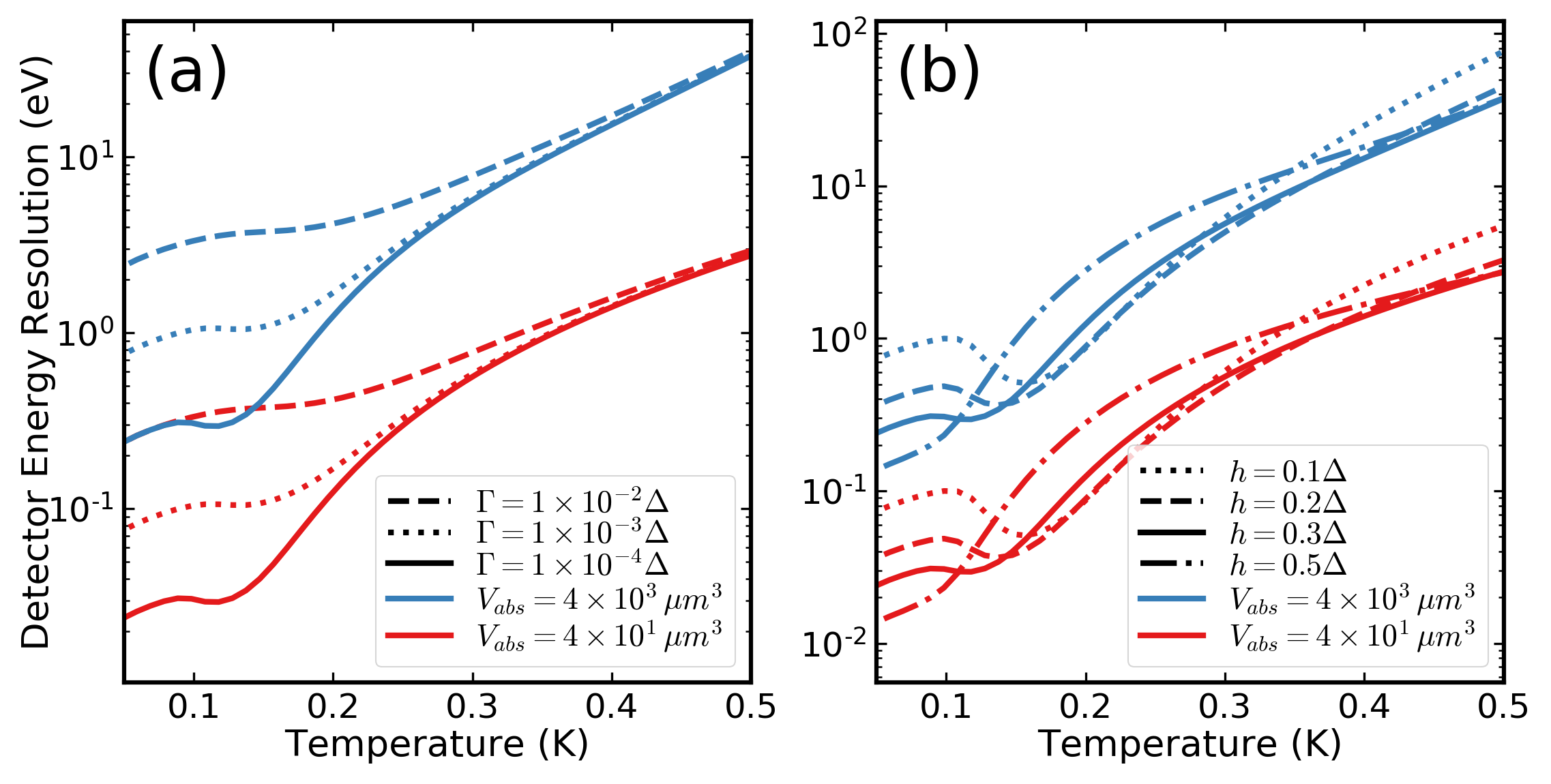}
        \caption{Numerical calculation of the detector energy resolution of SFTED with different broadening parameters ({\it left}) and exchange field ({\it right}). {\it Colored lines} indicate different absorber volumes. {\it Solid lines} are results using same parameters in Fig.~\ref{fig:fig2}. In these plots, the junction barrier was assumed to be AlO$_x$ with an area of $10^4\ \mu m^2$.}
        \label{fig:fig3}
    \end{center}
\end{figure}

The detector energy resolution ({\it solid lines} in Fig.~\ref{fig:fig2}) is proportional to the square root of the absorber volume, and decreases with decreasing temperature. However, the latter dependency becomes weaker at $T < 150 $ mK, and a saturation-like resolution floor appears. This saturation is caused by the sub-gap leakage current of the tunnel junction, described by a broadening parameter\cite{Dynes1978} $\Gamma$ (see Fig.~\ref{fig:fig3}(a)). We adopted a typical value for Al  \cite{panu2009} $\Gamma=10^{-4}\Delta$ in our calculations shown in Fig.~\ref{fig:fig2}, but it has been lowered to $10^{-7}\Delta$ with a multistage shielding\cite{Saira2012}, which would improve the ultimate resolution further.

The low temperature energy resolution limits are also very sensitive to the applied exchange field $h$, as shown in Fig.~\ref{fig:fig3}(b). A dramatic change in how the resolution depends on $h$ appears around $0.1\sim0.2$ K, and therefore, for each operation temperature of SFTED, a different optimal exchange field exists.

Looking back at the energy resolution results that include readout noise in Fig.~\ref{fig:fig2}, we see that a SQUID readout with $20\,fA/\sqrt{Hz}$ current noise matches perfectly with an SFTED ({\it dotted lines}). A SQUID with a superconducting flux transformer can in practice achieve a noise level of $<60\,fA/\sqrt{Hz}$ with an optimized design\cite{Drung2007,Drung2016} ({\it dash-dot}), and in that case SFTED will not be limited by readout noise at $T > 150\,mK$. An energy resolution less than $1\,eV$ with an absorber volume of $4\times10^3\,\mu m^3$ is thus predicted below 0.2 K.

\begin{figure}[htbp]
    \begin{center}
        \includegraphics[width=0.85\linewidth, keepaspectratio]{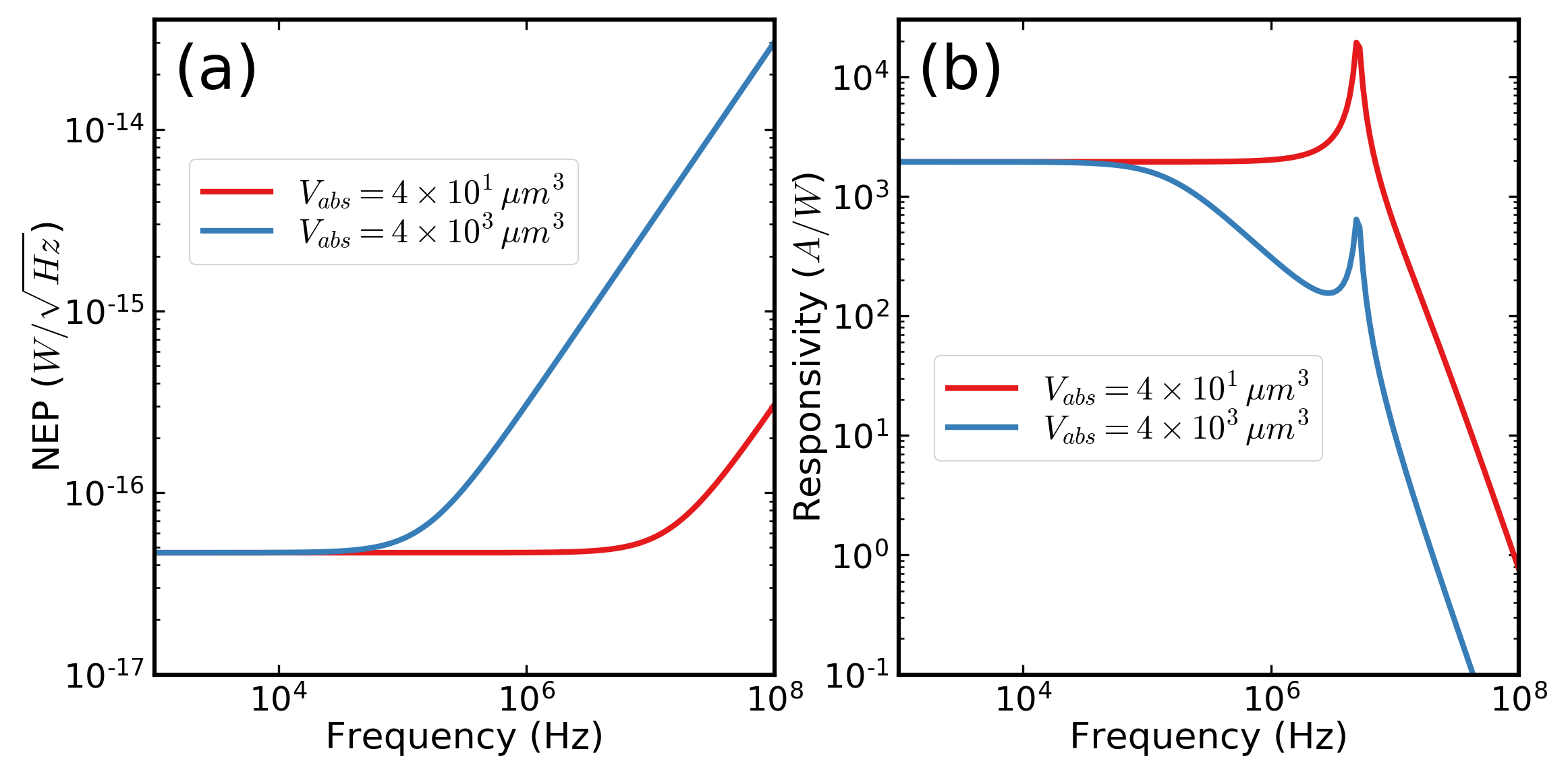}
        \caption{Numerical calculation of NEP ({\it left}) and responsivity ({\it right}) of SFTED with a flux transformer coupled SQUID readout. {\it Colored lines} indicate different absorber volumes. The junction barrier was assumed to be AlO$_x$ with area of $10^4\ \mu m^2$, with a broadening parameter of $\Gamma=10^{-4}\Delta$, and an exchange field of $h=0.3\Delta$.}
        \label{fig:fig4}
    \end{center}
\end{figure}

However, using a large flux transformer will introduce an electrical resonance into system and reduces the electrical bandwidth due to its large input inductance and parasitic capacitance. Fig.~\ref{fig:fig4} shows the frequency dependent NEP and responsivity of a SFTED, read out by a flux-transformer-coupled SQUID with an input inductance of $2\,\mu H$ and a total capacitance of $0.5\,nF$. It shows that detector bandwidth will be limited by the readout for small absorber volumes, while with the larger absorber, the roll-off of the thermal time constant dominates.

\section{Conclusions}
We have demonstrated that the novel superconductor-ferromagnet thermoelectric detector (SFTED) is a device that has promising characteristics when operated either as a bolometer or as a calorimeter. As a bolometer, SFTED with EuS as the tunnel barrier is competitive with the current state-of-the-art detector technologies. However, a current readout with a SQUID can be hard to achieve due to the high impedance of the EuS detector junction even with the highest ratio superconducting flux transformers. Other amplification techniques probing the voltage signal may thus be more suitable. 

On the other hand,  SFTED shows a promise as a calorimeter both in terms of energy resolution and detector bandwidth (speed), if used with an AlO$_x$ tunnel barrier, or if a low tunneling resistance EuS barrier can be made. A SQUID readout with a moderate flux-transformer-coupled input is feasible for this type of detector, without sacrificing much of the bandwidth. Combined with other beneficial features such as lack of bias and self-isolation by electron-phonon coupling, SFTED can be both attractive and practical for calorimetric applications that require large sensor arrays.

\begin{acknowledgements}
This study was supported by the Academy of Finland Project Number 298667 and by the European Union’s Horizon 2020 research 
and innovation programme under grant agreement No 800923 (SUPERTED). 
\end{acknowledgements}

\pagebreak

\end{document}